\documentclass[review,sort&compress,number]{elsarticle}

\usepackage[T1]{fontenc}
\usepackage[utf8]{inputenc}
\usepackage{lmodern}
\usepackage{microtype}
\usepackage[english]{babel}
\usepackage{enumitem}
\usepackage{amsmath}
\usepackage{amssymb}
\usepackage{graphicx}
\usepackage{wrapfig}
\usepackage{float}
\usepackage{xcolor}
\usepackage{algorithm}
\usepackage{algorithmic}
\usepackage{multirow}
\usepackage{bbm}
\usepackage{gensymb}
\usepackage{textcomp}
\usepackage{gensymb}
\usepackage{lineno}
\usepackage{graphicx}
\usepackage{wrapfig}
\usepackage{lscape}
\usepackage{rotating}
\usepackage{epstopdf}

\usepackage{fancyhdr}
\begin{document}
% MFDFA for increments
\begin{frontmatter}
\title{Linearity versus non-linearity in high frequency multilevel wind time series measured in urban areas}
%\author{Authors$^1,^2, ^3$}
\author{ Luciano Telesca$^{1}$, Mohamed Laib$^{2}$, Fabian Guignard$^2$, Dasaraden Mauree$^3$, Mikhail Kanevski$^2$}

\address{$^1$CNR, Istituto di Metodologie per l’Analisi Ambientale, Tito (PZ), Italy \\

$^2$IDYST, Faculty of Geosciences and Environment, University of Lausanne, Switzerland. \\
$^3$Solar Energy and Building Physics Laboratory, Ecole Polytechnique Fédérale de Lausanne, Switzerland \\

Corresponding author: Mohamed.Laib@unil.ch}

\begin{abstract}

In this paper, high frequency wind time series measured at different heights from the ground (from 5.5 to 25.5 meters) in an urban area were  investigated. The spectrum of each series is characterized by a power-law  behaviour at low frequency range, with a mean spectral exponent of about 1.5, which is rather consistent with the Kolmogorov spectrum of atmospheric turbulence. The detrended fluctuation analysis was applied on the magnitude and sign series of the increments of wind speed, in order to get information about the linear and nonlinear dynamics of the time series.  Both the sign series and magnitude series are characterized by two timescale ranges; in particular the scaling exponent of the magnitude series in the high timescale range seems to be related with the height of the sensor. This study aims to understand better high frequency wind speed in urban areas and to disclose the underlying mechanism governing the wind fluctuations at different heights.

\end{abstract}


\end{frontmatter}

%\linenumbers
\section{Introduction}
\label{intro}
Wind in urban environments, especially in built-up areas, is a crucial factor to consider in urban planning and layout. For instance, it was found that there could be significant over-speed of wind and vortices in connecting passage ways between two buildings. Also channelling effects of wind flow represent an environmental hazard for pedestrians \cite{Dutt1991}.  
To understand the impact of urban areas and/or buildings on wind and in order to improve the representation of land surface in evaluating building energy use \cite{Mauree2017aa, Mauree2018}, air pollutant dispersion, and renewable energy potential in urban planning scenarios \cite{Perera2018}, high quality wind data from experimental campaigns are essential. In general, monitored meteorological data, and wind in particular, are scarcely available with high vertical resolution and at a high frequency. Campaigns such as the BUBBLE \cite{Rotach2005} observation period provided useful information and data to develop and generalize new parameterization schemes. However, there is a strong need for such data and in multiple urban configurations to develop new approaches that can then be used in the evaluation of building energy use. The vertical profiles of wind speed, along with other environmental data, in the vicinity of buildings are crucial in the determination of the momentum and heat fluxes \cite{Mauree20177, Jarvi2018, Santiago2007}. Moreover, the high frequency acquisition of wind speed data is necessary for capturing small turbulent structures often present in an urban configuration \cite{Christen2009}. 

\begin{figure}
%%\rule{1cm}{1cm}width=\linewidth
\centering
\includegraphics[scale=0.5]{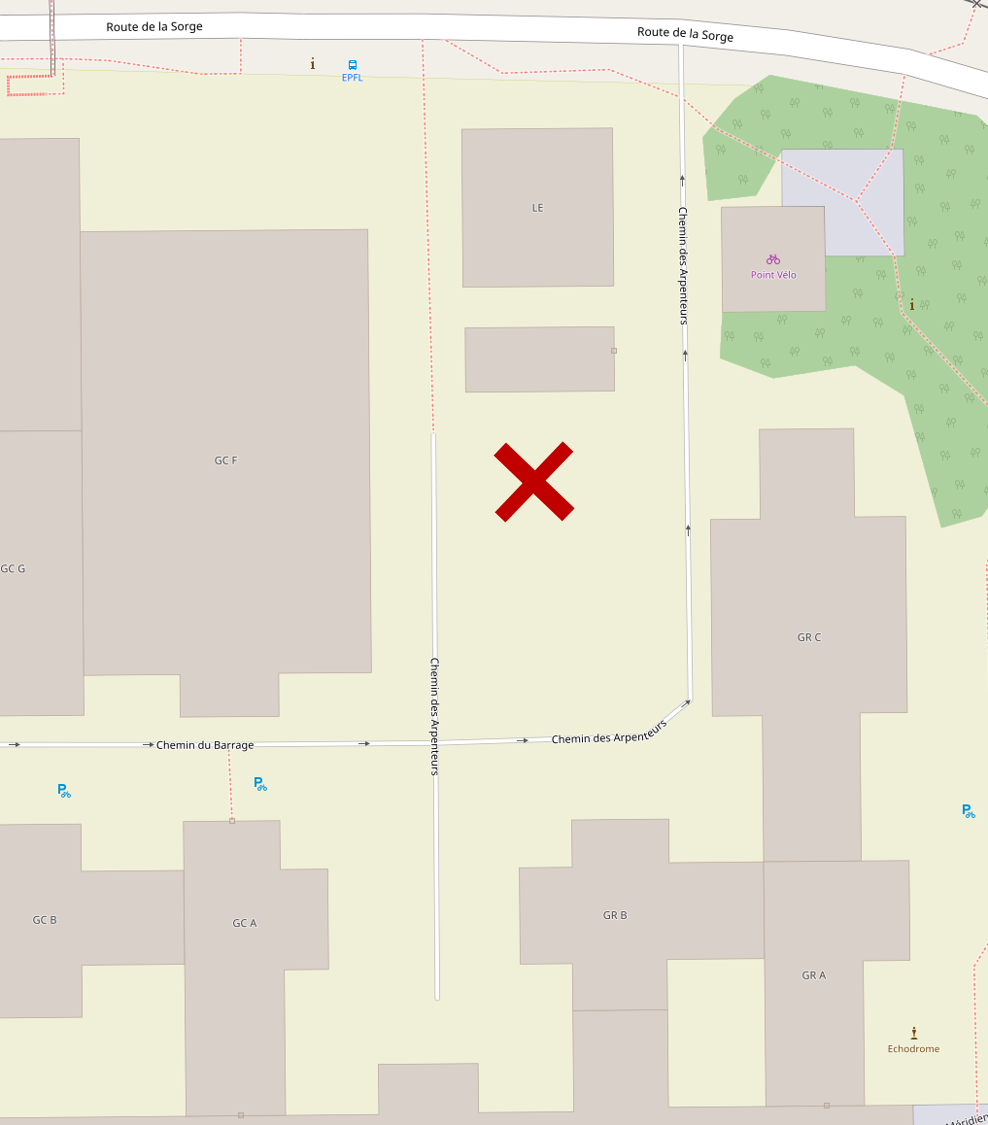}
\caption{Location of the setup on the EPFL campus (indicated with the red cross). This image is taken from Open Street Map whose copyright notices can be found here: https://www.openstreetmap.org/copyright (CC-BY-SA-2.0).}
\label{fig0}  
\end{figure}

Urban wind can be considered as affected by extremely complex interactions involving mean wind speed vertical gradient,  turbulence, shape, size, layout of buildings, etc. Such complexity of interactions features wind speed fluctuations as highly variable at any timescales, and its time dynamics is characterized by a non-linear behaviour. In order to disclose the non-linearity in wind speed series measured in urban environment and to understand its possible mechanisms and sources, a scaling analysis of multilevel wind speed time series was performed on data collected through an experiment implemented at Ecole Polytechnique Fédérale de Lausanne (EPFL), Switzerland (motus.epfl.ch). A 27 m high mast was installed in the campus, whose average building height is around 10 m, to measure with very high sampling frequency wind speed, along with other meteorological parameters \cite{Mauree2017bb,MAUREE2017325} (see Fig. \ref{fig0}). The experiment was carried out to understand how urban areas could impact on wind fluctuations.
The paper is organized as follows. First, the technical details of the experiment are briefly reported and exploratory and spectral analysis are performed on the data. Next, the methods employed in the data analysis (singular spectrum analysis, magnitude/sign decomposition and detrended fluctuation analysis) are described. Then, the results are presented and discussed, before the final remarks are summarized in the conclusions.

\section{Description of the experiment and data exploratory analysis}
\label{sec:1}
We analysed six wind speed time series measured from 28th November 2016 to 29th January 2017 by 3D sonic anemometers placed along the vertical axis of a 27 m high mast between 5.5 m and 25.5 m above the ground with 4 m spacing; the highest anemometer is  
sufficiently above the displacement height to be in a constant flux layer and thus in an undisturbed flow \cite{Rotach1999}. Data from the instruments are collected with a frequency of 20Hz based on the recommendations by Kaimal and Finnigan \cite{Kaimal1994} and are stored in a database at EPFL. Each anemometer acquires data for the three velocity components, the sonic speed and temperature and is stored in a text file with the corresponding time stamp. In this work, we focused on the one minute averages of wind speed.

\begin{sidewaysfigure}[ht]
%%\rule{1cm}{1cm}width=\linewidth
\centering
\includegraphics[width=\linewidth]{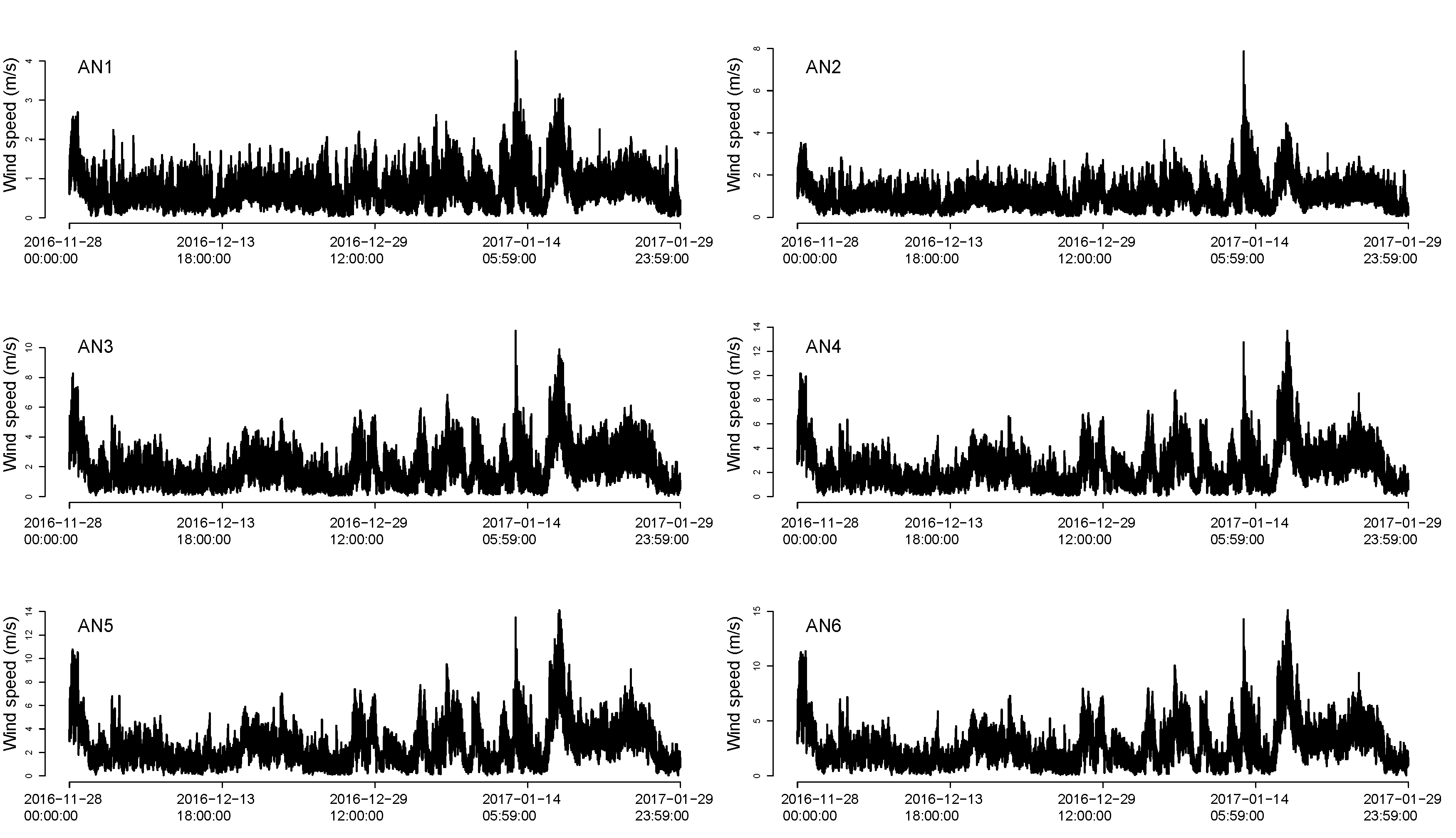}
\caption{Time series of wind speed recorded during the EPFL experiment.}
\label{fig1}  
\end{sidewaysfigure}

Hereafter, the series will be labelled as AN1, AN2, AN3, AN4, AN5, AN6, corresponding respectively to wind series recorded by anemometers situated at height 5.5 m, 9.5 m, 13.5 m, 17.5 m, 21.5 m, and 25.5 m.respectively. The data are shown in Fig \ref{fig1}.

Firstly, a distributional analysis has been performed fitting each wind series by three distributions that are generally used to describe wind data (Weibull, Gamma, and Generalized Extreme Value). Table \ref{Distr} shows the Kullback-Leibler divergence between the raw data and each distribution. The Kullback-Leibler divergence is used to evaluate the “similarity” between two distributions with density functions $p(x)$ and $q(x)$, respectively  \cite{kullback1951}:
%Given a random sample $X_1, \ldots, X_n$ from a probability distribution $P(x)$ with density function $p(x)$ over a non-negative support, assuming that the sample comes from a specific probability distribution $Q(x)$ with a density function $q(x)$, the KL divergence between $P(x)$ and $Q(x)$  is given by the following formula \cite{kullback1951}:

\begin{equation}
D_{KL}(p\|q)=\int_0^\infty p(x) \log \frac{p(x)}{q(x)}dx.
\end{equation}
The reader can refer to  \cite{Ebrahimi1992, waal1996} for more details.

The lowest value of the Kullback-Leibler divergence for the GEV distribution indicates that this would better describe the data. The supplementary file Fig1S.pdf shows the estimated distributions for each of the six wind series.

\begin{table}
\centering
\footnotesize
\begin{tabular}{r|rrr}
  \hline
 & Weibull & Gamma & GEV \\ 
  \hline

  AN1 & 4.08 & 2.49 & \textbf{2.20} \\ 
  AN2 & 6.77 & 3.92 & \textbf{3.07} \\ 
  AN3 & 2.82 & 2.21 & \textbf{1.81} \\ 
  AN4 & 2.62 & 2.19 & \textbf{1.75} \\ 
  AN5 & 2.41 & 2.07 & \textbf{1.68} \\ 
  AN6 & 2.45 & 2.09 & \textbf{1.70} \\ 
   \hline
\end{tabular}
\caption{The Kullback-Leibler divergence of the three distributions for each wind series.}
\label{Distr}
\end{table}

\begin{figure}
%%\rule{1cm}{1cm}width=\linewidth
\centering
\includegraphics[width=\linewidth]{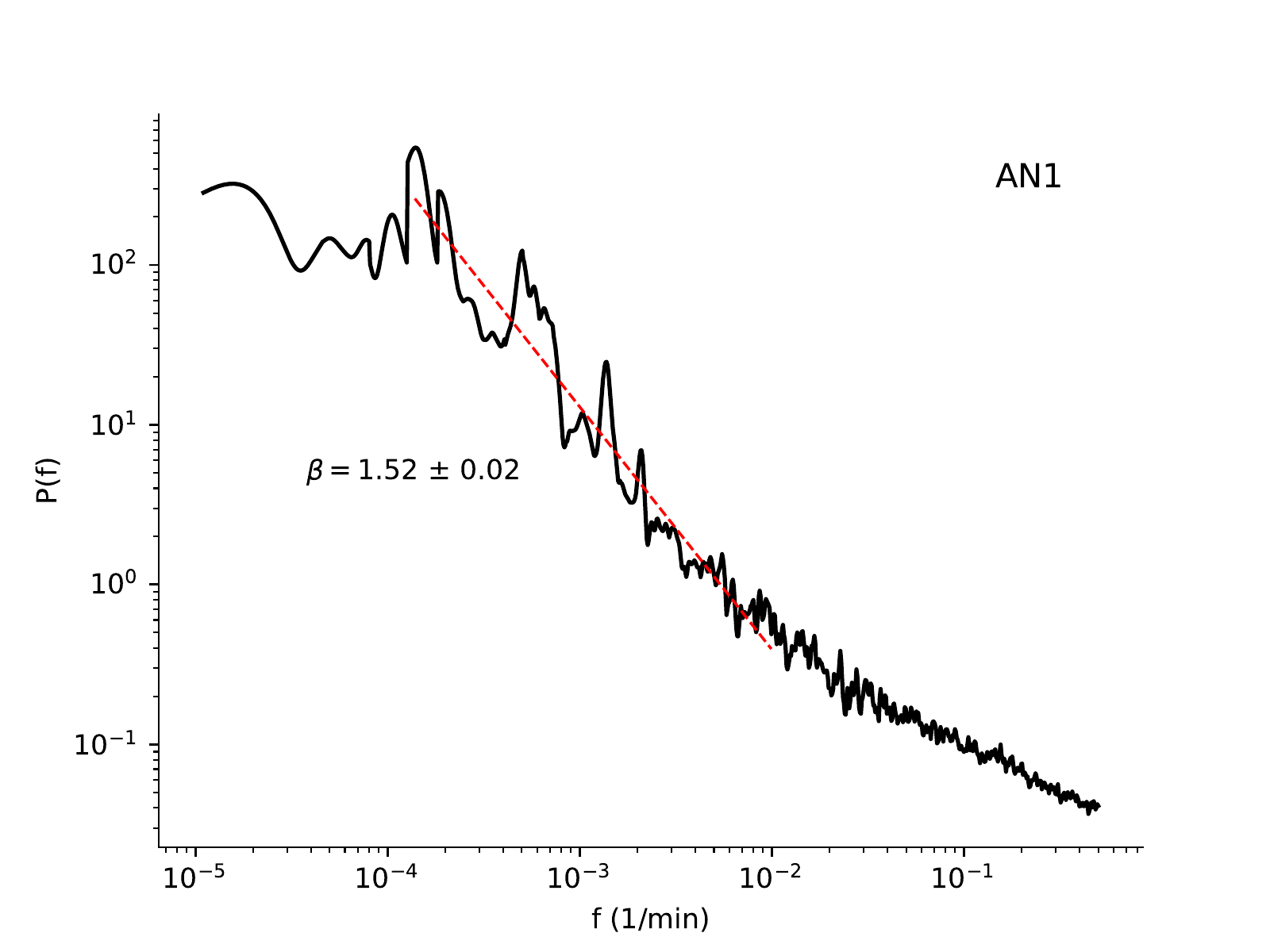}
\caption{Power spectral density of AN1 series.}
\label{fig2}  
\end{figure}

The spectral content of each series was examined by calculating the power spectral density. We used the Logarithmic frequency axis Power Spectral Density (LPSD) method \cite{Trobs2006}, which is well suited for long time series and computes the optimal frequency resolution individually for each Fourier frequency on a logarithmic frequency axis. For the mathematical details on the LPSD method, the reader can refer to\cite{Trobs2006}. Fig \ref{fig2} shows, as an example, the power spectral density of AN1, plotted in log-log scales. The supplementary figure Fig2S.pdf shows the power spectral densities of all series. Fitting the power spectral density by a least square line between about  $10^{-4}$ and $10^{-2}$ $min^{-1}$ , the slope of the line is the spectral exponent $\beta$ that ranges around $1.5$; this value is in close agreement with the values of spectral exponent for wind series estimated by \cite{Li2018}, and consistent with the Kolmogorov spectrum \cite{Kolmogorov1941} of atmospheric turbulence \cite{Bandi2017}.

\begin{table}
\centering 
\footnotesize
\begin{tabular}{l|cccccc}
\hline
 & AN1 & AN2  & AN3  & AN4  & AN5  & AN6 \\
\hline
 AN1 & $1$ & $0.92$  & $0.82$  & $0.82$  & $0.81$  & $0.81$ \\
 AN2 & $0.92$ &  $1$  & $0.86$  & $0.83$  & $0.83$   & $0.83$  \\
 AN3 & $0.82$ & $0.86$  & $1$  & $0.96$  & $0.97$  & $0.98$ \\
 AN4 & $0.82$ & $0.83$  & $0.96$  & $1$  & $0.99$  & $0.98$ \\
 AN5 & $0.81$ &  $0.83$   & $0.97$  & $0.99$  & $1$  & $0.99$ \\
 AN6 & $0.81$ &  $0.83$   & $0.98$  & $0.98$  & $0.99$  & $1$ \\
\hline

\end{tabular}
\caption{Pearson’s correlation coefficient of any two wind series.}
\label{T1}
\end{table}

\begin{table}
\centering 
\footnotesize
\begin{tabular}{l|cccccc}
\hline
 & AN1 & AN2  & AN3  & AN4  & AN5  & AN6 \\
\hline
AN1 & $1$ & $0.99$  & $0.93$  & $0.93$  & $0.93$  & $0.94$ \\
AN2 & $0.99$ &  $1$  & $0.93$  & $0.92$  & $0.93$   & $0.93$  \\
AN3 & $0.93$ & $0.93$  & $1$  & $1$  & $0.99$  & $0.99$ \\
AN4 & $0.93$ & $0.92$  & $1$  & $1$  & $1$  & $1$ \\
AN5 & $0.93$ &  $0.93$   & $0.99$  & $1$  & $1$  & $1$ \\
AN6 & $0.94$ &  $0.93$   & $0.99$  & $1$  & $1$  & $1$ \\
\hline

\end{tabular}
\caption{Pearson’s correlation coefficient of any two trends.}
\label{T2}
\end{table}

\section{Methods}
\subsection{The singular spectrum analysis method}
The singular spectrum analysis (SSA) \cite{Vautard1989} is a well developed and powerful tool for time series analysis. It is a decomposition method, in which data are represented as the superposition of independent components. These components
are generally trend, periodic/quasi-periodic oscillations, and structureless noise  \cite{Hassani2007}.

Let $X_N=(x_1, \ldots, x_N)$ be a time series of length $N$, $M$ $(1<M<N)$ is an integer called window length.

The SSA decomposition can be illustrated in two steps \cite{GOLYANDINA2014}:
\begin{enumerate}
\item Embedding  of the original time series into a sequence of lagged vectors of size $M$ by forming $K=N-M+1$ lagged vectors:

$X_i=(x_i, \ldots, x_{i+M-1})^T$, $i=1, \ldots, K$.

Then the trajectory matrix of the series $X_N$ is given by

$\,$

$\textbf{X}= [X_1 : \ldots : X_k]= (x_{ij})^{M,K}_{i,j=1}
= \left(
  \begin{array}{ c c c c c}
     X_1 & X_2 & X_3 & \ldots & X_k \\
     X_2 & X_3 & X_4 & \ldots & X_{k+1} \\
     X_3 & X_4 & X_5 & \ldots & X_{k+2}  \\
      \ldots &  \ldots &  \ldots & \ddots &  \ldots \\
     X_M & X_{M+1} & X_{M+2} & \ldots & X_N 
  \end{array} \right)$
  
Both rows and columns are subseries of the original series. The trajectory matrix $\textbf{X}$ is Hankel since it has equal elements on anti diagonal.

\item Let $\{ P_i \}^M_i=1$ be an orthonormal basis in $R^M$. Consider the following decomposition of $\textbf{X}$
\begin{equation}
\label{grpss}
\textbf{X} = \sum_{i=1}^M P_iQ_i^T= \textbf{X}_1+ \ldots+\textbf{X}_M
\end{equation}
where $Q_i= \textbf{X}^TP_i$, and $\lambda_i=\Vert \textbf{X}_i\Vert^2_\tau$.

$\{ P_i \}^M_{i=1}$ are the eigenvectors ordered in the decreasing order of the eigenvalues, of the Toeplitz lagged correlation matrix 
 $\textbf{C}$ whose entries are:

\begin{equation}
c_{ij}=\frac{1}{N-|i-j|}\sum_{m=1}^{N-|i-j|}x_mx_{m+|i-j|}, 1\leqslant i, j\leqslant M.
\end{equation}
%The Toeplitz case is suitable o
\end{enumerate}
For more details on the use of the SSA and its extension, the reader can refer to \cite{GOLYANDINA2014, Korobeynikov2010, Golyandina2015}.
 
 The SSA is based on the calculation of the Toeplitz lagged correlation matrix of the series, which depends on the number M of independent components of the time series M. The choice of the lag M is an important step in SSA and should result by a trade-off between quantity of information (given by larger M) and degree of statistical confidence (smaller M) \cite{Ghil2002}. If the time series is characterised by a cycle with a period of T, the SSA will not be able to identify it from other long-term fluctuations unless M is longer than T. In \cite{Vautard1992}, it is suggested to keep M lower than $\frac{1}{N}$, where N is the length of the time series. With that rule of thumb, SSA can identify about $\frac{M}{10}$ significant components before oscillations start to be lumped together. Therefore, the lag M should be larger than the longest periodicity under study but in the same time smaller than $\frac{N}{5}$ and larger than $\frac{M}{10}$ times the number of significant components that might be present. Khan and Poskitt suggested that $M=(logN)^c,$ $1.5\leq c \leq 2.5$ \cite{Khan2010}.

After, the eigenvalues $\lambda_k$ and eigenvectors $E_{kj}$ of the Toeplitz lagged correlation matrix are computed. Then, the constructed components $r_{ik}$ of the time series $y_i$ can be obtained by
\begin{equation}
r_{ik}=\frac{1}{M}\sum_{j=1}^M a_{i-j,k}E_{jk}, \; M \leq i \leq N-M +1,
\end{equation}
where $a_{ik}$ is the $k^{th}$ principal component given by
\begin{equation}
a_{ik}=\sum_{j=1}^M y_{i+j}E_{jk}, \; 0 \leq i \leq N-M.
\end{equation}

The eigenvalue $\lambda_k$ indicates the fraction of the total variance of the series in the $k^{th}$ component. Sorting the eigenvalues in decreasing order corresponds to reconstructing the components by decreasing information about the original time series \cite{Schoellhamer2001}. In general, the first reconstructed component contains most of the variance and presents the trend while the other ones are oscillations and structureless noise.

\subsection{Detrended fluctuation analysis}
Detrended fluctuation analysis (DFA) is considered as one of the effective alternative methods to the power spectral density in the identification of scaling behaviour in non-stationary time series. It was introduced in \cite{Peng1994} and is widely used to detect long-range correlation proprieties of time series. DFA can be used to extract information about the type of temporal fluctuations in data.  The DFA can briefly by described as follows:

\begin{enumerate}
\item Let $x_i$ be a time series with a total number of samples $N$ ($i = i, \ldots, N$). The time series $x_i$ is integrated as
\begin{equation}
y(k) = \sum_{i=1}^{k} x_i-\bar{x}
\end{equation}
where $\bar{x}$ is the mean value of the time series.

\item The obtained time series $y_k$ is divided into boxes of equal length $n$.
\item For each $n$-size box, $y_k$ is fitted using a polynomial $y_{n,k}$, which represents the local trend of the box.
\item $y_k$ is detrended by subtracting the local trend $y_{n,k}$, then the root-mean fluctuation for each box is computed by 
\begin{equation}
F(n)=\sqrt{\frac{1}{N} \sum_{k=1}^{N} [y_k - y_{n,k}]^2}.
\end{equation}
\item The above procedure is repeated for all n-size boxes to provide a relationship between $F(n)$ and the box size n, which for long-range power law correlated signals is as a power-law
\begin{equation}
F(n) \propto n^\alpha
\end{equation}

\item The scaling exponent $\alpha$ quantifies the strength of the long-range power-law correlation of the time series $x_i$. If $\alpha=0.5$, the series is uncorrelated; if $\alpha>0.5$ the correlations of the series are persistent;
%where persistence means that a large (small) value (compared to the average) is more likely to be followed by a large (small) value
if $\alpha<0.5$, the correlations of the series are antipersistent, which indicates that a large (small) value is more likely to be followed by a small (large) value comparing to the average. 

\item Depending on the degree of the detrending polynomial in the above step 3, we define DFA-1, DFA-2,$\ldots$, if the degree of the detrending polynomial is respectively 1, 2, $\ldots$

\end{enumerate} 
%is able to identify and quantify scaling in noisy signals for a wide range of correlations \cite{Hu2001}

\subsection{Magnitude and sign decomposition}

It was recently shown that fluctuations of time series can be featured by two components, the magnitude (absolute value) and the sign (direction) components \cite{Ashkenazy2001}. These two components would reveal  the inner interactions of a system, whose resulting force would determine the magnitude and the direction of the fluctuations.

There is no relationship between the scaling in time series and scaling in their magnitudes. Some time series are uncorrelated while their magnitudes are correlated, like certain econometric time series \cite{Liu1999}. A recent test for non-linearity \cite{Schreiber2000} was based on the evaluation of the scaling properties of a time series. Thus, different systems can display similar scaling laws although their non-linear properties could be different. 

Ashkenazy et al. \cite{Ashkenazy2003b} proposed to analyse the long-range correlated (scaling) time series, by decomposing them  in two sub-series, the magnitude and sign of the increments \cite{Ashkenazy2001}:
\begin{equation}
\Delta x_i = sgn(\Delta x_i) |\Delta x_i|
\end{equation}
where $\Delta x_i$ is the increment of the time series $x_i$, $sgn(\Delta x_i)$ are the sign of the increments and $|\Delta x_i|$ their magnitude; for $x=0$ we define $sgn(0)=0$ \cite{Ashkenazy2003b}.

Analysing by the DFA the magnitude and sign series, as defined above, information about the existence of long-range correlated structures in the analysed series can be obtained. Correlation in the magnitude series suggests that an increase with a certain magnitude is more probably followed by an increase with similar magnitude. Anticorrelation in the sign suggests that a positive increment would more likely be followed by a negative increment and vice versa. It was shown by Ashekazy et al.  \cite{Ashkenazy2003b} that scaling analysis of magnitude series can furnish information about nonlinearity in the original series, while that of the sign series relates mainly to linear properties.

\section{Results and discussion}

In this paper, six wind time series of one minute averages recorded at the EPFL at heights above the ground from 5.5 m to 25.5 m were analysed (see Fig.\ref{fig1}). The different measuring heights from the ground were motivated by the aim to take into account two main characteristics of the experiment: the canopying effect, revealed by the lowest anemometers and the undisturbed flow revealed by the highest ones. It can be clearly seen that all the six time series seem to be characterized by similar variability, although the amplitude of the variation increases with the height from the ground.  The close similarity among the wind series is confirmed by the Pearson correlation analysis (Table \ref{T1}), which shows a correlation coefficient ranging between $0.81$ and $0.99$, indicating a high shape similarity. Such very close similarity among the series indicates that a common forcing should govern the main variability of the wind at different heights.

\begin{figure}
%%\rule{1cm}{1cm}width=\linewidth
\centering
\includegraphics[width=\linewidth]{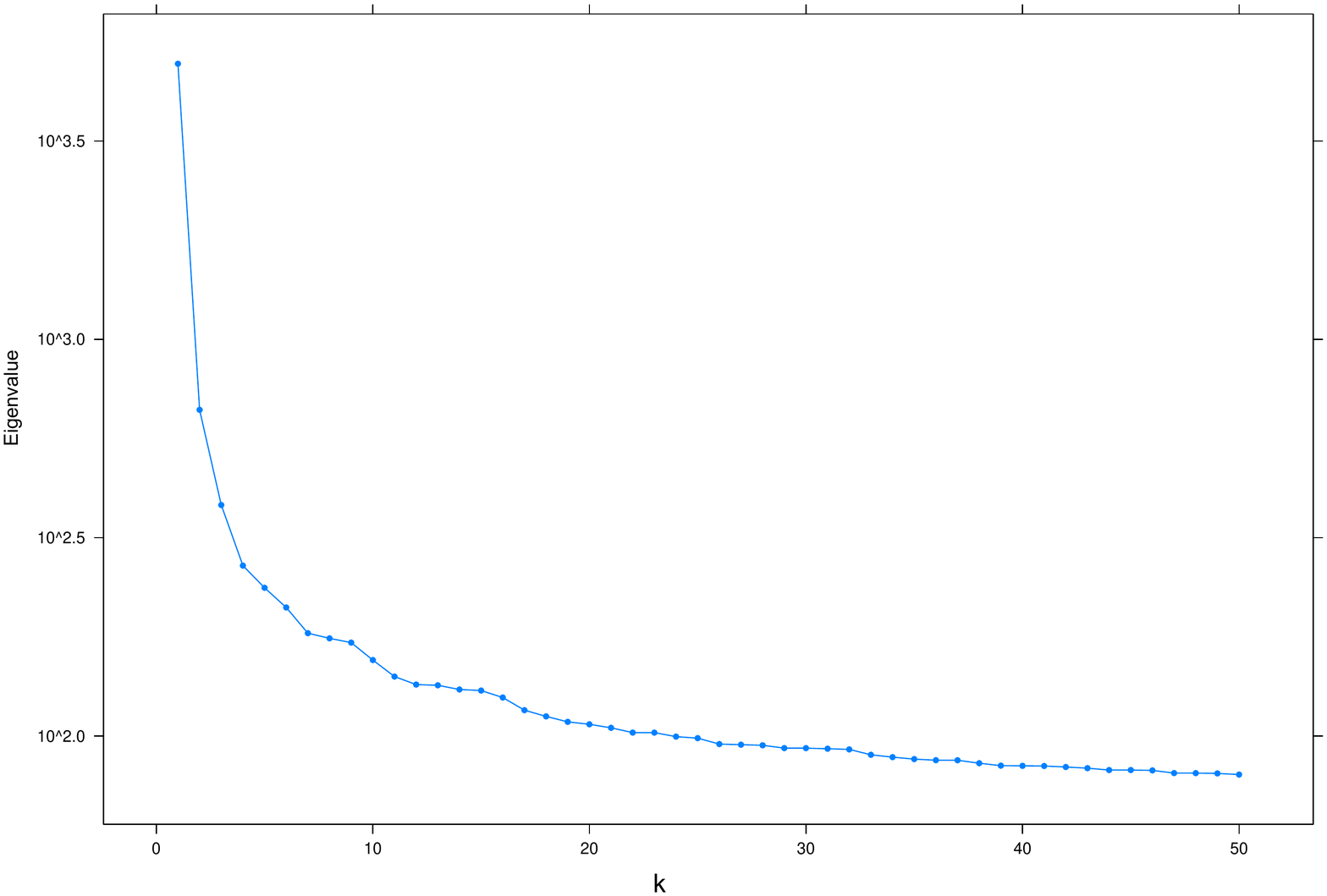}
\caption{Eigenvalues spectrum of the first 50 components of AN1.}
\label{fig3}  
\end{figure}

\begin{sidewaysfigure}[ht]
%%\rule{1cm}{1cm}width=\linewidth
\centering
\includegraphics[width=\linewidth]{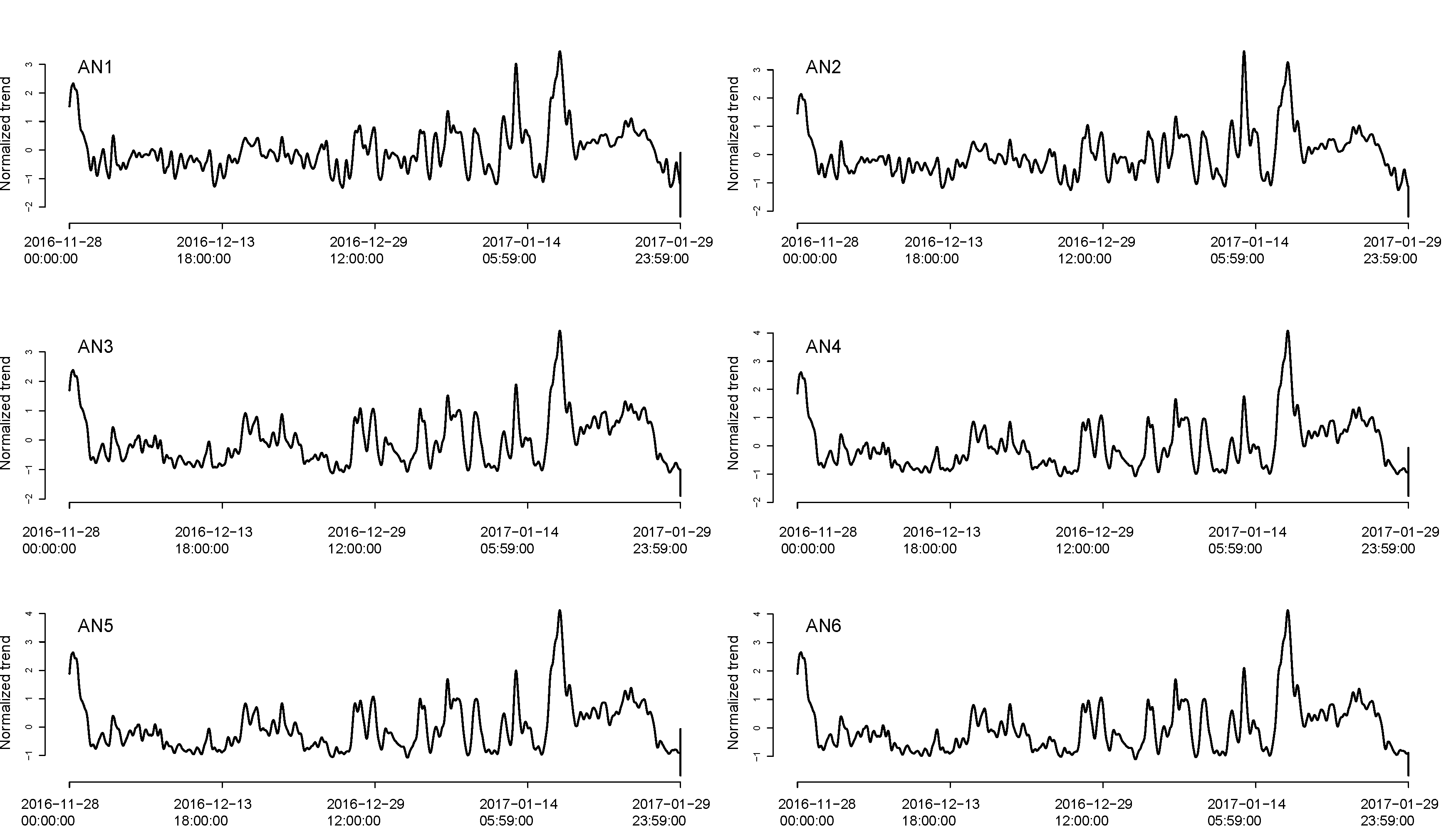}
\caption{Trend obtained by SSA method.}
\label{fig4}  
\end{sidewaysfigure}

In order to extract such common forcing, we applied SSA to each time series. On the base of the Khan and Poskitt’s \cite{Khan2010} criterion the window $M=(logN)^{2.5}$ was chosen; thus each time series was decomposed into $M=440$ independent components. Fig \ref{fig3} shows the eigenvalue spectrum of the series AN1, as an example. The supplementary figure Fig3S.pdf shows the eigenvalue spectra of all series. The eigenvalue spectrum furnishes a measure of the contribution of each component to the total variance of the series. The first eigenvalue corresponds to the component that contributes maximally to the total variance of the series and generally coincides with the trend. Fig. \ref{fig4} shows synoptically the trends of all the wind series.

\begin{sidewaysfigure}[ht]
%%\rule{1cm}{1cm}width=\linewidth
\centering
\includegraphics[width=\linewidth]{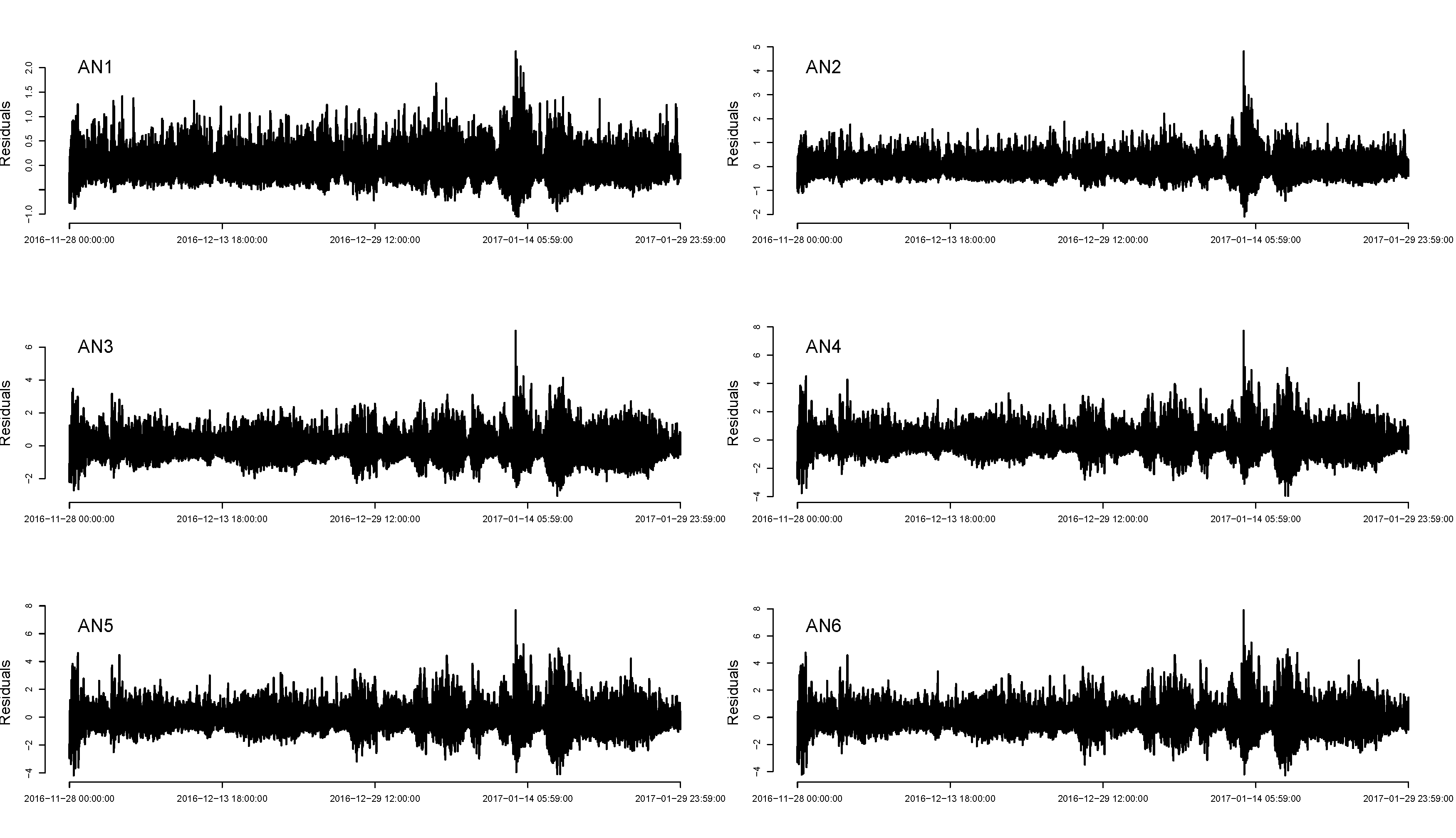}
\caption{Residuals data after removing the trend using the SSA method.}
\label{fig5}  
\end{sidewaysfigure}

\begin{sidewaysfigure}[ht]
%%\rule{1cm}{1cm}width=\linewidth
\centering
\includegraphics[width=\linewidth]{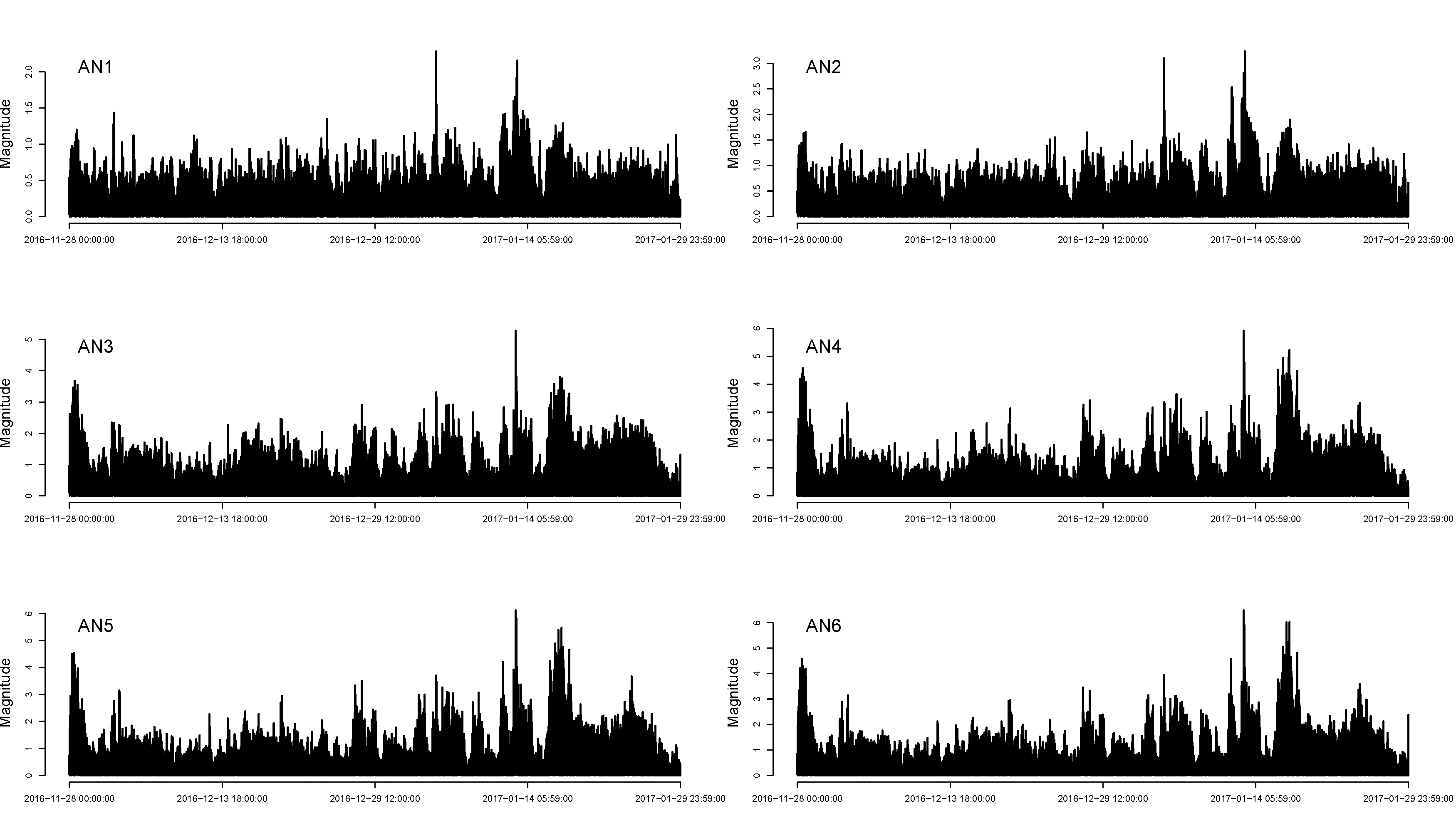}
\caption{Magnitude series.}
\label{fig6}  
\end{sidewaysfigure}

As it can be clearly seen, the trends behave very similarly (as confirmed by  the Pearson correlation analysis (Table \ref{T2})) and could be considered as a forcing that could be the main responsible of the shape similarity among the original wind series. Therefore, in order to characterize the inner time dynamics of the wind series, we removed from each time series its trend, and investigated the residual series. Fig \ref{fig5} shows the residuals of the six wind series. To each residual series the magnitude and sign decomposition method was applied. Thus, after calculating the increments, their absolute values and signs were obtained. Fig. \ref{fig6} shows the magnitude and sign of the increments of the residuals of all series.

The magnitude and sign decomposition allow us to investigate the linear/non-linear properties of the residual wind speed series. To this aim, the DFA-1 was applied to each magnitude and sign series as defined above for timescales ranging between 10 min and 1/10 of the length of the series. Fig. \ref{fig7} shows the fluctuation curve of the magnitude series of the increments of all the residuals, while Fig. \ref{fig8} shows that of the sign series. In both series, but more evident in the magnitudes, two timescale ranges are visible, one involving the lower timescales from $10$ min to $\sim70$ min and the other involving the higher timescales from $\sim300$ min to $\sim9070$ min. The slopes of the line fitting by the least squares method the fluctuation function in each timescale region represent the scaling exponents in that region. The scaling exponent $\alpha_{mag}$ of the magnitude series in the low and high timescale regions for all the series is shown in  Fig. \ref{fig9}, while that of the sign series $\alpha_{sign}$ is shown in Fig. \ref{fig10}.
Using DFA-2, the values of the scaling exponent slightly changes in the low timescale range for the magnitudes, but it is almost the same in the high timescale range, indicating that the first degree of the detrending polynomial is rather sufficient to remove the existing trends. 

The scaling exponent of the magnitude series is larger in the high timescale range than in the low one, indicating that the magnitude series is  more correlated in the high timescale range than in the low one, where it tends to be more randomly distributed. The sign series reflects the same behavior of the magnitude series, being characterized by a scaling exponent larger in the high timescale range than in the low one. In the low timescale range, however, the sign series exhibits anticorrelated behavior. In the high timescale range the scaling exponent of the magnitude series tends to saturate for heights above 13.5 m. Therefore, the rate of variability of the scaling exponent with the height is relatively large below 13.5 m, while it becomes relatively very small for height above 13.5 m. This crossover height could signal a sort of change of the mechanisms governing wind time dynamics; below this crossover height canopying phenomena could emerge for the specific layout of buildings, and micro-turbulence effects could affect the increase of rate of variability of the scaling exponent; while above it, wind is free to flow, and the possible absence of obstacles makes the wind to have a scaling behaviour that is less dependent on the height.    

 However, a common feature that characterizes all the magnitude series is that the value of the scaling exponent is significantly different from 0.5,  indicating that non-linearity strongly characterizes the wind series at any height.

To strengthen our findings, a surrogate data analysis was performed generating randomized Fourier-phase series with the same power spectrum and histogram of the increment series; in this case, only the non-linear properties are eliminated, since the power spectrum and the histogram are kept. The randomized Fourier-phase surrogates were generated as described in \cite{Schreiber2000, Ashkenazy2003b}.  One hundred surrogates were generated. Each surrogate was decomposed in magnitude and sign series, before applying  DFA-1. Fig \ref{fig12} shows the mean and the standard deviation of the scaling exponents of the magnitude and sign series of the randomized Fourier-phase surrogates calculated in the high timescale range. It is clearly observed that the sign series derived from the surrogates are characterized by a scaling behaviour rather identical to the scaling of the sign series derived from the original series. On the contrary, the magnitude series derived from the surrogates are characterized by a scaling exponent significantly different from that of the original magnitude series; furthermore, they show a uncorrelated behaviour, clearly different from the strongly correlated behavior exhibited by the original magnitude series. 
These results suggest that the magnitude series conveys information about the nonlinear properties of the wind series, while the linear properties are mainly described by the sign series. 

\section{Conclusion}
In this work, the minute averages of wind speed measured at different heights from the ground in an urban settlement were investigated. The spectral characteristics of all the series are rather consistent with those of the Kolmogorov spectrum of atmospheric turbulence. The performed analysis of linearity/nonlinearity by using the magnitude/sign decomposition method combined with the detrended fluctuation analysis has permitted to identify the crossover height of about 13.5 m that separates two possible different mechanisms of generation of wind fluctuations; below this crossover height canopying phenomena joint to micro-turbulence effects would dominate, while above it the impact of building layout would be much lower. This study contributes to a better comprehension of high frequency wind speed fluctuations in urban areas and could be helpful to a better urban planning.

\section{Acknowledgements}
This research was partly supported by the National Research Programme 75
"Big Data" (PNR75) of the Swiss National Science Foundation (SNSF).

M. Laib thanks the support of "Société Académique Vaudoise" (SAV) and the Swiss Government Excellence Scholarships. 

L. Telesca thanks the support of the "Scientific Exchanges" project n° 180296 funded by the SNSF.

%\nocite{*}

\bibliography{xampl}
\bibliographystyle{elsarticle-num}

\newpage

\begin{sidewaysfigure}[ht]
%%\rule{1cm}{1cm}width=\linewidth
\centering
\includegraphics[width=\linewidth]{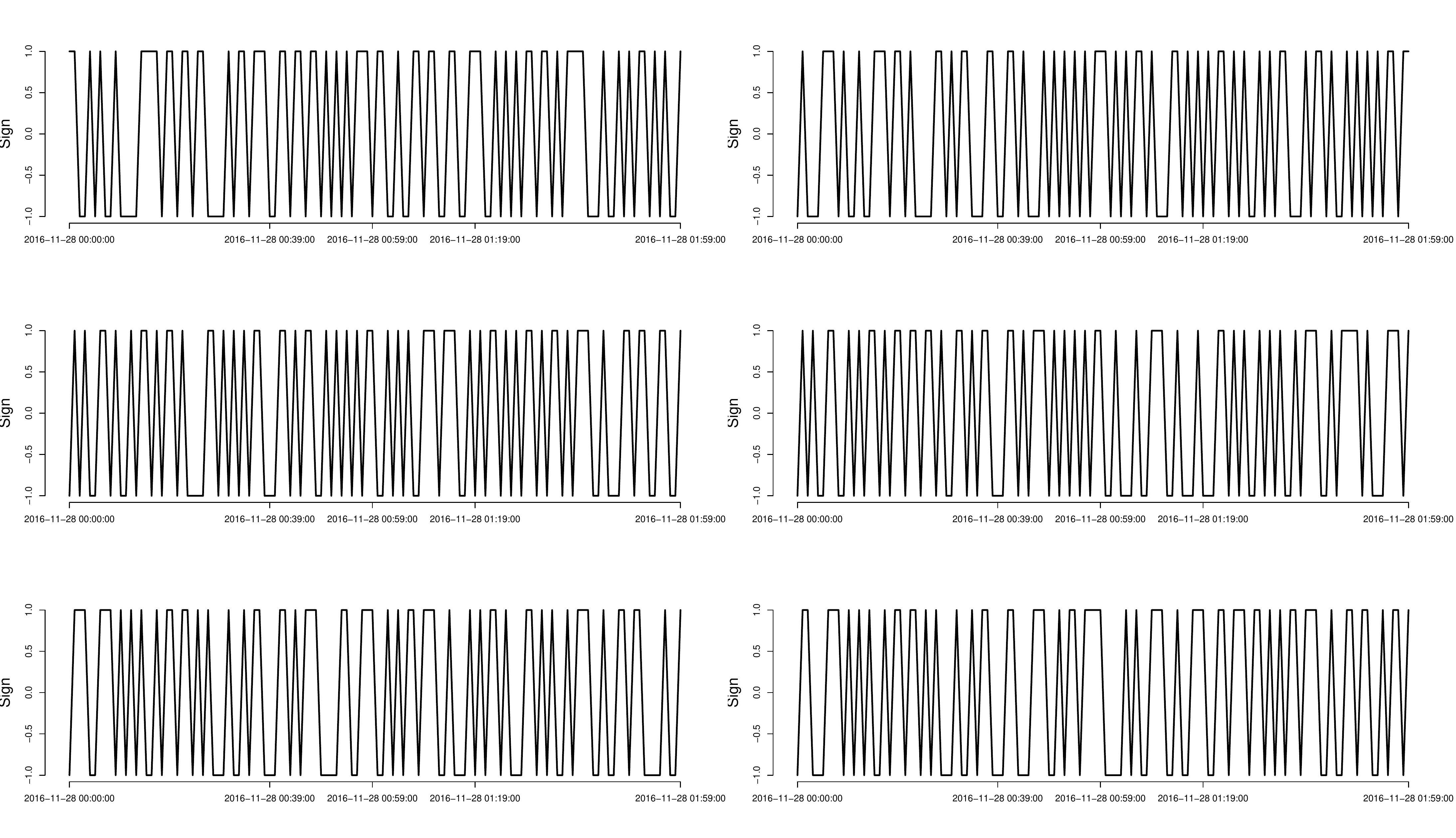}
\caption{Excerpt of 2 hours of sign series.}
\label{fig6b}  
\end{sidewaysfigure}

\begin{figure}
%%\rule{1cm}{1cm}width=\linewidth
\centering
\includegraphics[width=\linewidth]{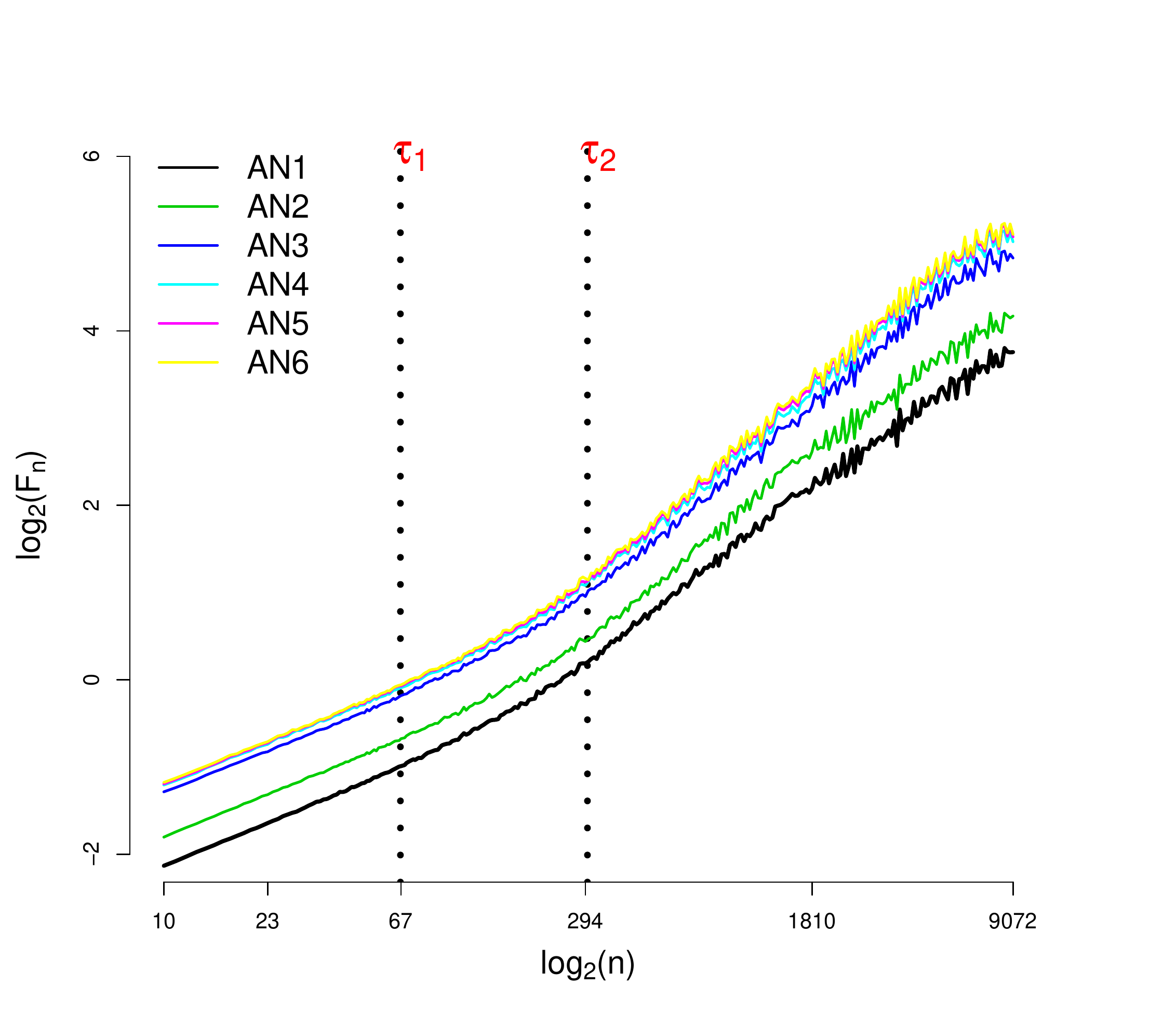}
\caption{Fluctuation function of the magnitude series.}
\label{fig7}  
\end{figure}

\begin{figure}
%%\rule{1cm}{1cm}width=\linewidth
\centering
\includegraphics[width=\linewidth]{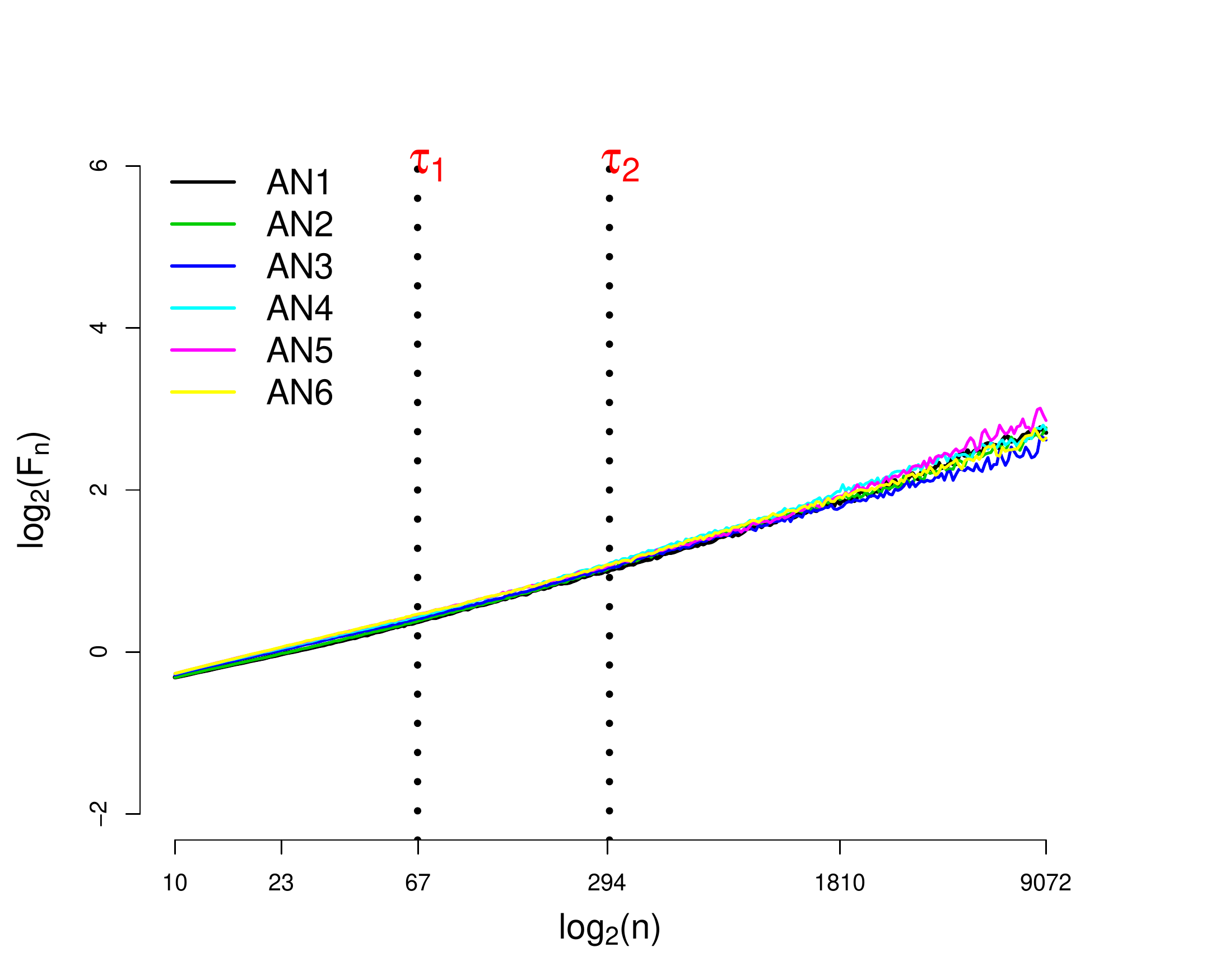}
\caption{Fluctuation function of the sign series.}
\label{fig8}  
\end{figure}

\begin{figure}
%%\rule{1cm}{1cm}width=\linewidth
\centering
\includegraphics[width=\linewidth]{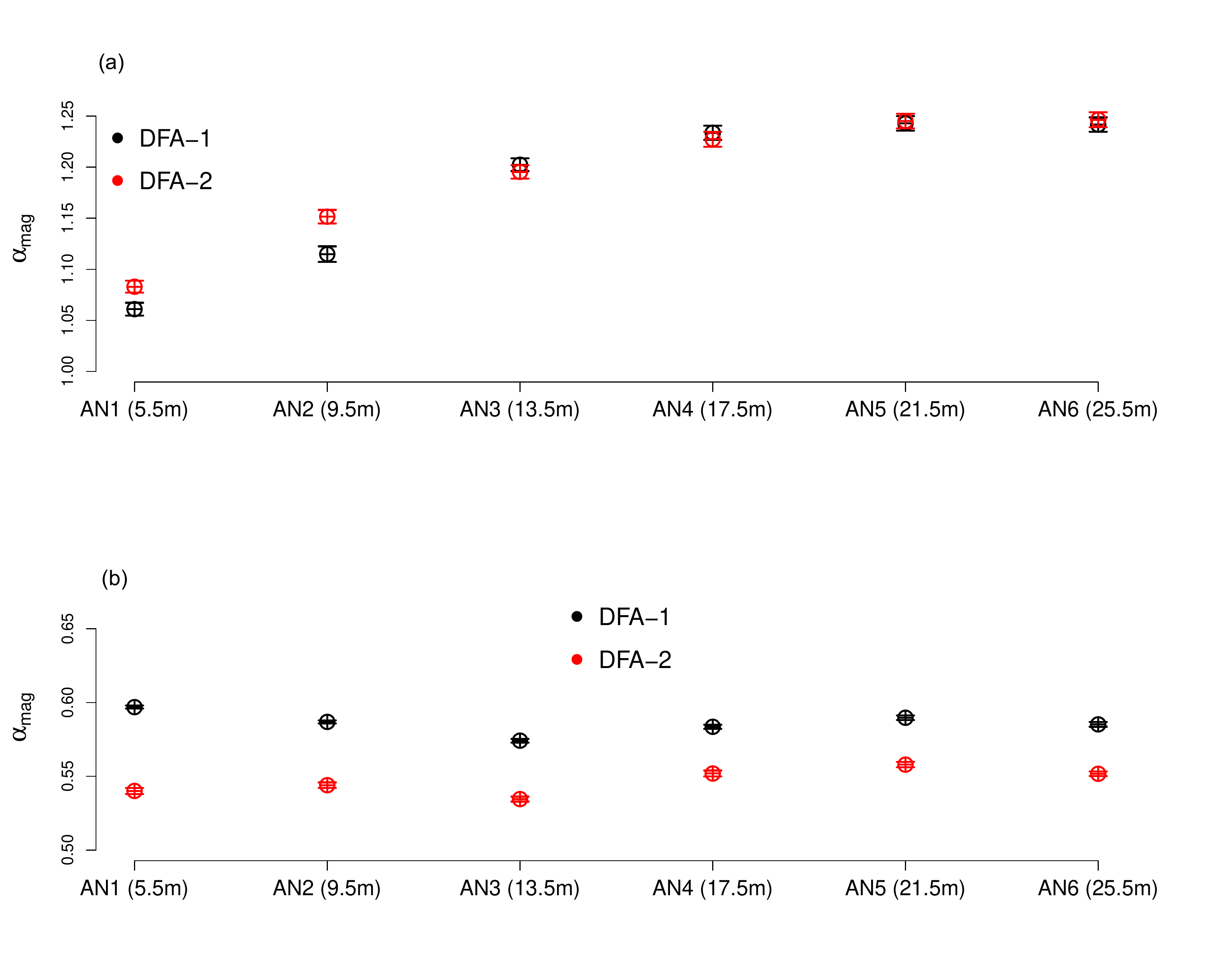}
\caption{$\alpha_{mag}$ in the high timescale range (a) and low timescale range (b).}
\label{fig9}  
\end{figure}

\begin{figure}
%%\rule{1cm}{1cm}width=\linewidth
\centering
\includegraphics[width=\linewidth]{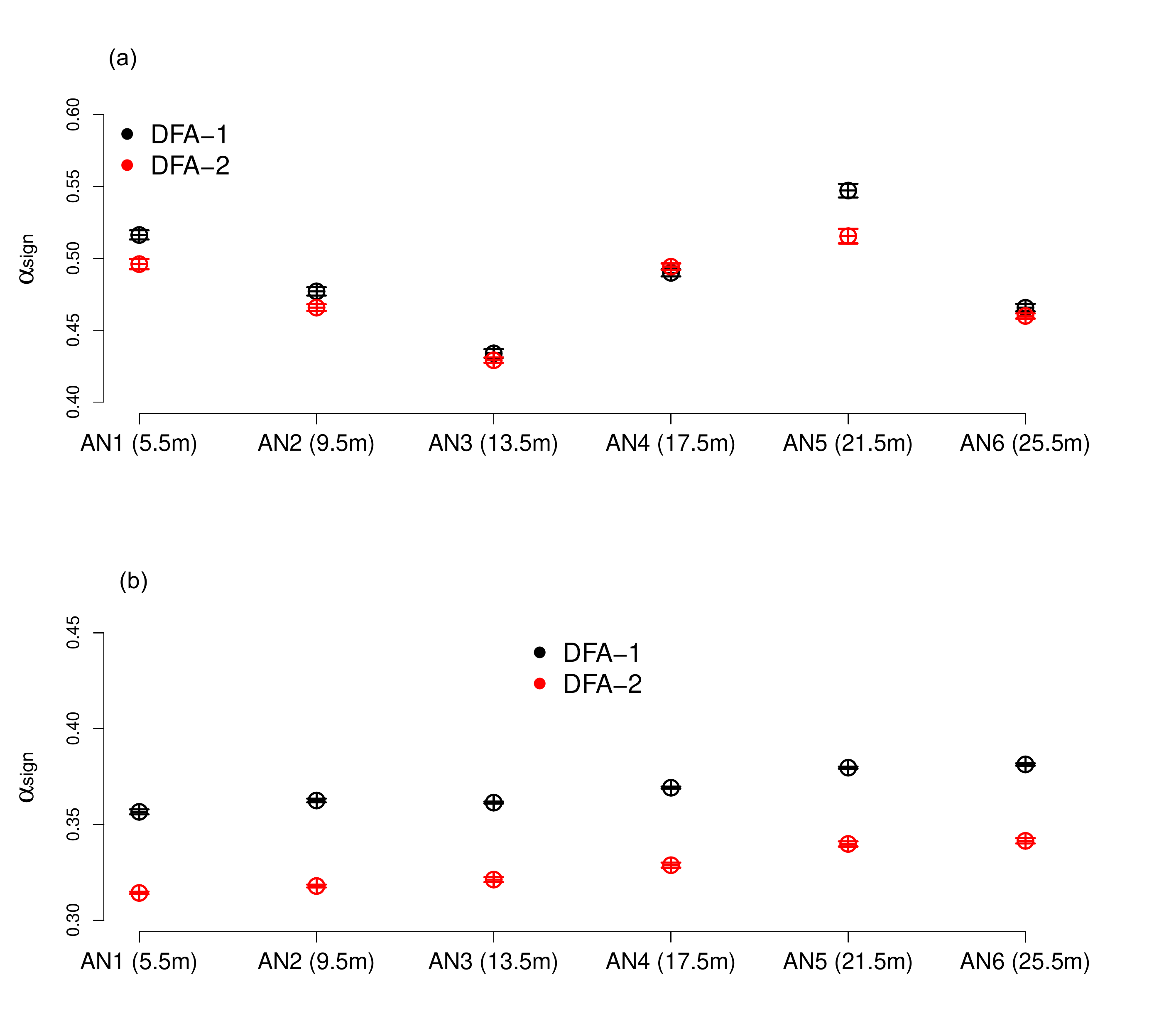}
\caption{$\alpha_{sign}$ in the high timescale range (a) and low timescale range (b).}
\label{fig10}  
\end{figure}

\begin{figure}
%%\rule{1cm}{1cm}width=\linewidth
\centering
\includegraphics[width=\linewidth]{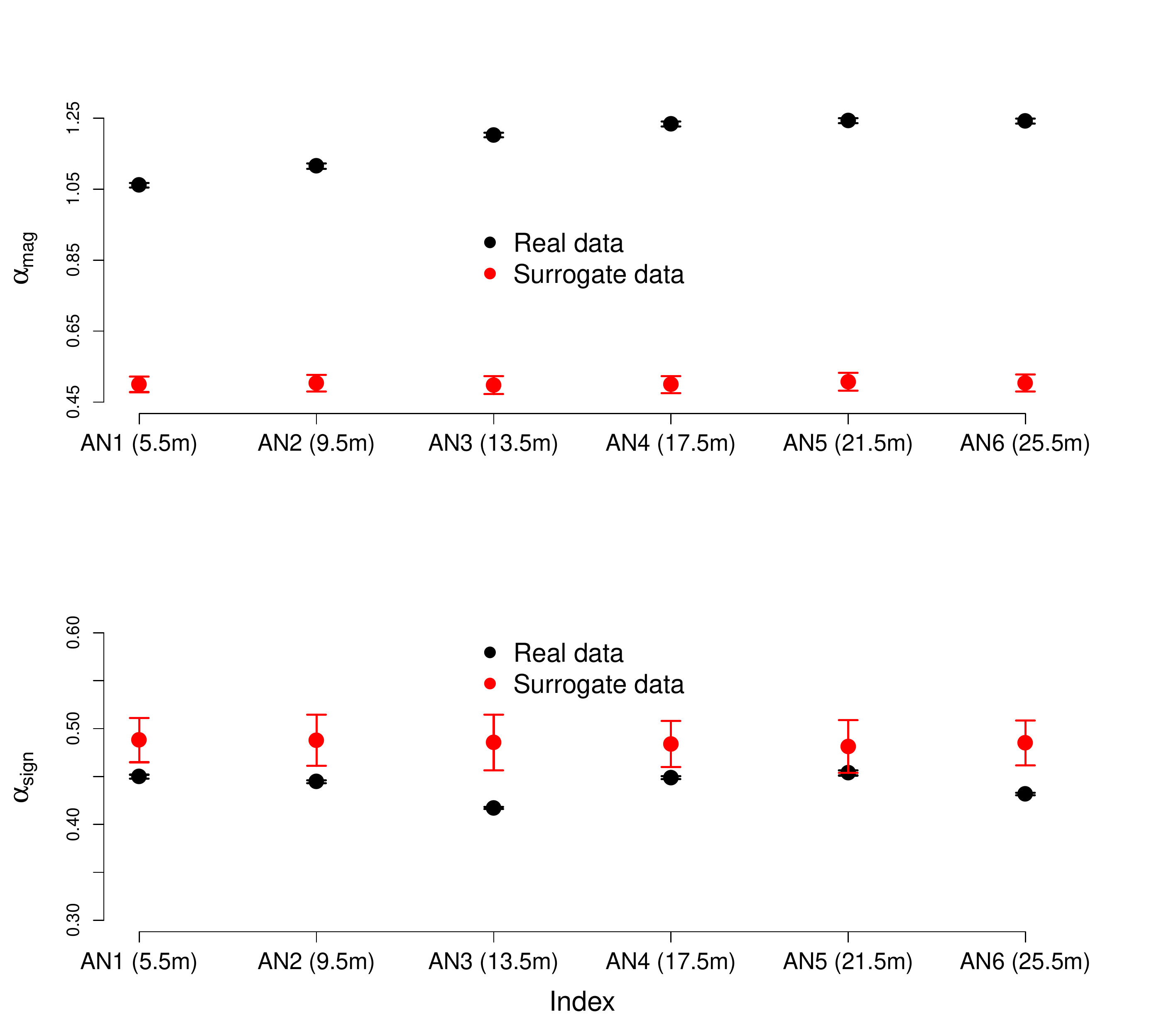}
\caption{ Comparison between the scaling exponents for the magnitude series (a) and the sign series (b) of the real (black) and surrogate series (red).}
\label{fig12}  
\end{figure}

%
%\begin{figure}
%%%\rule{1cm}{1cm}width=\linewidth
%\centering
%\includegraphics[width=\linewidth]{figs/surrogate_sign.pdf}
%%%\includegraphics[width=\linewidth]{stations_b.pdf}
%\caption{Scaling exponent for the sign series (real data versus surrogate data).}
%\label{fig15}  
%\end{figure}

\end{document}